\documentclass[sigconf,natbib=true]{acmart}
\AtBeginDocument{%
  }

\renewcommand\footnotetextcopyrightpermission[1]{}
\usepackage{hyperref}
\usepackage{booktabs}
\usepackage{multirow}
\usepackage{algorithm}
\usepackage{algorithmic}
\usepackage{enumitem}
\usepackage{enumerate}
\usepackage{xcolor}
\usepackage{graphicx}
\usepackage{subcaption}
\usepackage{diagbox} 
\usepackage{makecell}
\usepackage{array}
\usepackage{colortbl}
\usepackage{pifont}
\usepackage{amsmath}
\definecolor{colorAdam}{RGB}{48, 135, 170}
\definecolor{colorMuon}{RGB}{232, 75, 58}

\newcommand{\eg}{\emph{e.g.}}
\newcommand{\ie}{\emph{i.e.}}

\begin{document}

\title{MuonRec: Shifting the Optimizer Paradigm Beyond Adam in Scalable Generative Recommendation}

\author{Rong Shan}
\email{shanrong@sjtu.edu.cn}
\authornote{Equal contribution.}
\affiliation{
  \institution{Shanghai Jiao Tong University}
  \city{Shanghai}
  \country{China}
}

\author{Aofan Yu}
\email{yu_aofan@sjtu.edu.cn}
\authornotemark[1]
\affiliation{
  \institution{Shanghai Jiao Tong University}
  \city{Shanghai}
  \country{China}
}

\author{Bo Chen}
\email{renze03@kuaishou.com}
\affiliation{
  \institution{Kuaishou Technology}
  \city{Beijing}
  \country{China}
}

\author{Kuo Cai}
\email{caikuo@kuaishou.com}
\affiliation{
  \institution{Kuaishou Technology}
  \city{Beijing}
  \country{China}
}

\author{Qiang Luo}
\email{luoqiang@kuaishou.com}
\affiliation{
  \institution{Kuaishou Technology}
  \city{Beijing}
  \country{China}
}

\author{Ruiming Tang}
\email{tangruiming@kuaishou.com}
\affiliation{
  \institution{Kuaishou Technology}
  \city{Beijing}
  \country{China}
}

\author{Han Li}
\email{lihan08@kuaishou.com}
\affiliation{
  \institution{Kuaishou Technology}
  \city{Beijing}
  \country{China}
}

\author{Weiwen Liu}
\email{wwliu@sjtu.edu.cn}
\affiliation{
  \institution{Shanghai Jiao Tong University}
  \city{Shanghai}
  \country{China}
}

\author{Weinan Zhang}
\email{wnzhang@sjtu.edu.cn}
\affiliation{
  \institution{Shanghai Jiao Tong University}
  \city{Shanghai}
  \country{China}
}

\author{Jianghao Lin}
\email{linjianghao@sjtu.edu.cn}
\authornote{Corresponding author.}
\affiliation{
  \institution{Shanghai Jiao Tong University}
  \city{Shanghai}
  \country{China}
}

\renewcommand{\shortauthors}{Rong Shan et al.}

\begin{abstract}

Recommender systems (RecSys) are increasingly emphasizing \textit{scaling}, leveraging larger architectures and more interaction data to improve personalization. Yet, despite the optimizer’s pivotal role in training, modern RecSys pipelines almost universally default to Adam/AdamW, with limited scrutiny of whether these choices are truly optimal for recommendation. In this work, we revisit optimizer design for scalable recommendation and introduce \textbf{\textit{MuonRec}}, the first framework that brings the recently proposed Muon optimizer to RecSys training. Muon performs orthogonalized momentum updates for 2D weight matrices via Newton--Schulz iteration, promoting diverse update directions and improving optimization efficiency. We develop an open-sourced training recipe to recommendation models and evaluate it across both traditional sequential recommenders and modern generative recommenders. Extensive experiments demonstrate that MuonRec reduces converged training steps by an average of \textbf{32.4\%} while simultaneously improving final ranking quality. Specifically, MuonRec yields consistent relative gains in NDCG@10, averaging \textbf{12.6\%} across all settings, with particularly pronounced improvements in generative recommendation models. These results consistently outperform strong Adam/AdamW baselines, positioning Muon as a promising new optimizer standard for  RecSys training. Our code is available~\footnote{\url{https://anonymous.4open.science/r/MuonRec-E447}}.


\end{abstract}



\maketitle

\section{Introduction}

Recommender systems (RecSys) play an indispensable role in modern digital platforms by personalizing user experiences and driving key business objectives~\cite{li2023large, lin2023can}. A prominent trend in both RecSys research and industrial deployment is the emphasis on \textbf{\textit{scaling}}, which involves expanding model architectures and training datasets to unprecedented sizes~\cite{deng2025onerec, zhu2025rankmixer, zhai2024actions, chen2024hllm}. Representative models such as OneRec~\cite{deng2025onerec} and RankMixer~\cite{zhu2025rankmixer} have surpassed 1B parameters, representing a thousand-fold increase in model capacity compared to traditional architectures, which typically operate within a range of 0.1M to 100M parameters. 

Within the paradigm of extreme scaling, the optimizer acts as the critical bridge that translates structural capacity into predictive power. Despite its pivotal role, however, the choice of optimizer remains a largely unexamined default in modern recommender systems, often eclipsed by the rapid evolution of model architectures. Current practices almost universally default to adaptive optimizers like Adam~\cite{kingma2014adam} and AdamW~\cite{loshchilov2017decoupled_adamw}, typically without rigorously evaluating their suitability for the unique characteristics of recommendation tasks, particularly for large-scale models that must sustain high-frequency updates on web-scale interaction data~\cite{he2020lightgcn, deng2025onerec, zhou2018deep}.

Recently, the Muon optimizer~\cite{liu2025muon, jordan2024muon} has emerged as a promising alternative to Adam for large language model (LLM) training~\cite{liu2025muon, si2025adamuon, shah2025practical}. Its novel approach optimizes $2D$ parameters using orthogonalized gradient momentum based on Newton-Schulz iteration~\cite{stotsky2019unified, stotsky2022systematic}. Intuitively, this mechanism promotes diversity in weight updates, preventing the model from converging along only a few dominant directions. In large-scale language modeling, Muon has demonstrated remarkable success, achieving performance comparable to or better than AdamW with significantly fewer training FLOPs, thereby exhibiting high training efficiency~\cite{liu2025muon, shah2025practical, si2025adamuon, tveit2025muon}. Nevertheless, it remains an open question whether Muon can deliver similar gains in efficacy and training efficiency within recommender systems. Specifically, would it enable large-scale recommendation models to process more interaction data within the same training timeframe, while maintaining the stability required for high-frequency updates? Exploring this is especially vital for web-scale systems where training is both resource- and data-intensive.

\begin{figure}[t]
  \centering
  \includegraphics[width=\linewidth]{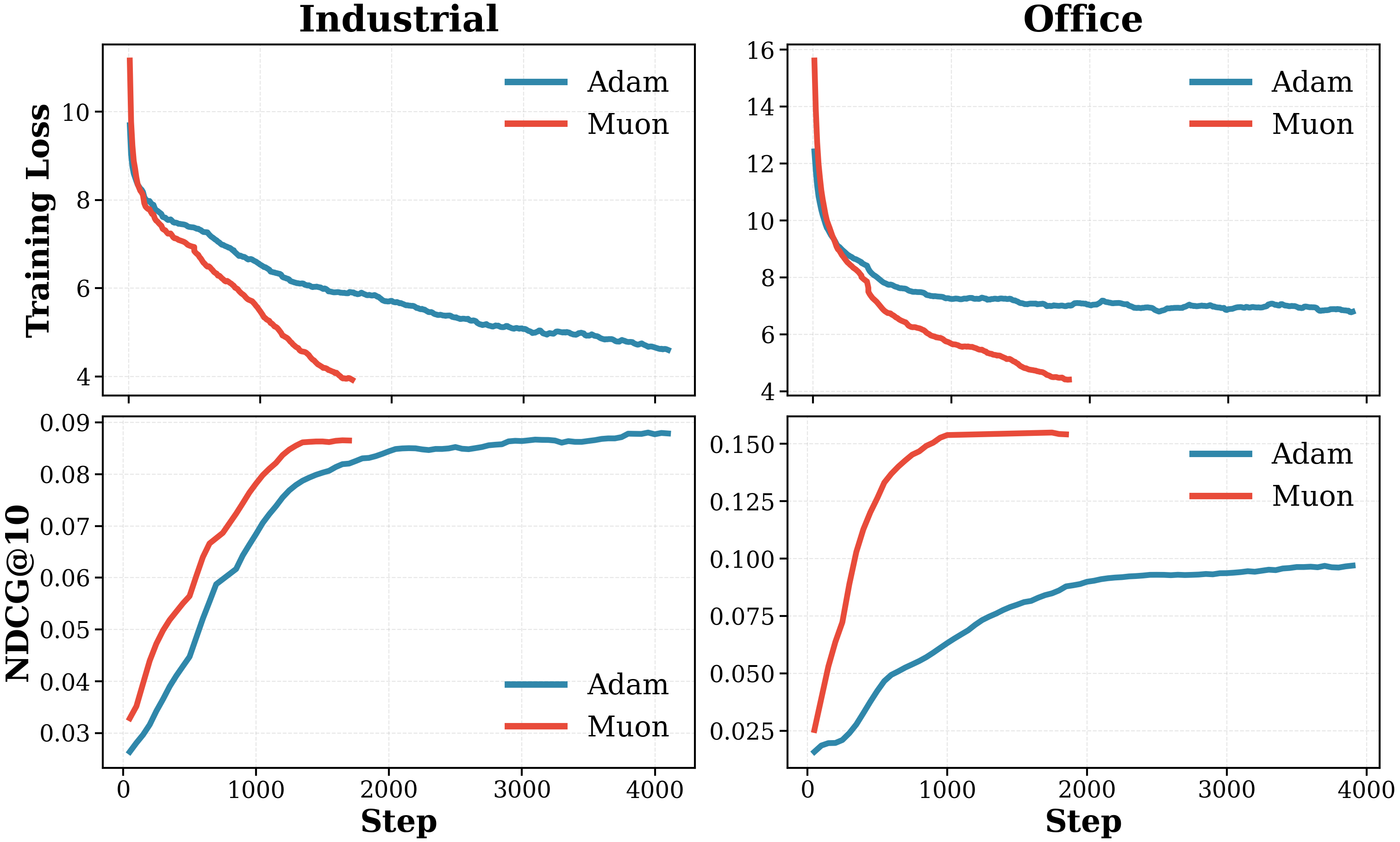}
  \vspace{-0.6cm}
  \caption{Training dynamics of \textbf{TIGER}~\cite{rajput2024recommender} on \textit{Industrial} (left) and \textit{Office} (right) datasets. The top row displays training loss, while the bottom row reports the NDCG@10 curves. The \textbf{\textcolor{colorMuon}{Muon}} optimizer demonstrates significantly faster convergence and superior final performance compared to \textbf{\textcolor{colorAdam}{Adam}}.}
  \vspace{-0.5cm}
  \Description{Training dynamics of the \textbf{TIGER} model on \textbf{Industrial} (top row) and \textbf{Office} (bottom row) datasets. The left column displays the Training Loss, while the right column reports the NDCG@10 curves. The \textbf{Muon} optimizer (red) demonstrates significantly faster convergence and superior final performance compared to \textbf{Adam} (blue), validating its efficiency and effectiveness for generative recommendation.}
  \label{fig:tiger_dynamics}
\end{figure}

To this end, we introduce \textbf{\textit{MuonRec}}, the first framework to adopt the Muon optimizer for recommendation. Through comprehensive experiments on both traditional sequential models and modern generative recommendation models, we demonstrate that Muon consistently enables more efficient training and leads to superior final performance compared to Adam (see Figure~\ref{fig:tiger_dynamics} for example). Our results provide the first empirical evidence and practical insights, establishing Muon as a promising new standard for optimizing modern large-scale recommender systems.

Our core contributions can be listed as follows:
\begin{itemize}[leftmargin=10pt]
    \item  We introduce MuonRec, a framework that brings the Muon optimizer to recommender model training. To the best of our knowledge, this is the first systematic study that adapts Muon to recommendation and provides an open-sourced training recipe for both sequential and generative recommendation models.
    \item We demonstrate through extensive experiments on various types types and sizes of recommendation models demonstrate that Muon consistently outperforms widely used adaptive optimizers (\ie, Adam), achieving better final recommendation quality and more efficient training under the same compute budget.
\end{itemize}
We believe MuonRec not only demonstrates the feasibility of adopting Muon for recommender model training, but also uncovers a previously overlooked dimension in the scaling laws of RecSys. Beyond data and model size, MuonRec opens a new dimension of matrix-structured optimization for the RecSys community, suggesting that revisiting fundamental optimizer design is key to further scaling recommendation models.

\section{MuonRec Framework}

\subsection{Problem Formulation}
Let $\mathcal{U}$ and $\mathcal{V}$ denote the sets of users and items, respectively.
For each user $u \in \mathcal{U}$, a chronological interaction sequence
$H_u = [v_i]_{i=1}^{t}$ is observed, where $v_i \in \mathcal{V}$ is the $i$-th interacted item and $t$ is the sequence length.
The objective of sequential recommendation is to estimate a conditional distribution $P(v_{t+1}\mid H_u)$ and predict the most probable next item~\cite{boka2024survey, fang2020deep}.

Conventional \textit{sequential recommendation models} represent each item $v\in\mathcal{V}$ with an integer ID and directly model $P(v_{t+1}\mid H_u)$.
In contrast, \textit{generative recommendation} (GenRec) represents each item by a hierarchical \textit{Semantic ID} (SID) tuple $s_v=[c_j]_{j=1}^{l}$, where $c_j$ is chosen from the $j$-th codebook. Accordingly, next-item prediction is cast as autoregressively generating $s_{v_{t+1}}$ conditioned on $H_u$, enabling an LLM-style Transformer backbone~\cite{li2025survey, deng2025onerec,hou2025generative, vaswani2017attention}.

\subsection{Adam Optimizer}
Adam~\cite{kingma2014adam} maintains exponential moving averages of the first and second moments of gradients.
Given parameters $\theta$ and stochastic gradient $g_t=\nabla_\theta \mathcal{L}_t(\theta_{t-1})$ at the $t$-th iteration, Adam updates the parameter $\theta$ by the following equations:
\begin{align}
\mathbf{m}_t &= \beta_1 \mathbf{m}_{t-1} + (1-\beta_1)\mathbf{g}_t, \\
\mathbf{v}_t &= \beta_2 \mathbf{v}_{t-1} + (1-\beta_2)\mathbf{g}_t\odot \mathbf{g}_t, \\
\hat{\mathbf{m}}_t &= \frac{\mathbf{m}_t}{1-\beta_1^t},\quad
\hat{\mathbf{v}}_t = \frac{\mathbf{v}_t}{1-\beta_2^t}, \\
\theta_t &= \theta_{t-1} - \eta \frac{\hat{\mathbf{m}}_t}{\sqrt{\hat{\mathbf{v}}_t}+\epsilon},
\end{align}
where $\odot$ is the elementwise product, $\beta_1,\beta_2\in[0,1)$, $\eta$ is the learning rate, and $\epsilon$ is a small constant.
AdamW~\cite{loshchilov2017decoupled_adamw} further applies \textit{decoupled} weight decay:
\begin{equation}
\theta_t = \theta_{t-1} - \eta\left(\frac{\hat{\mathbf{m}}_t}{\sqrt{\hat{\mathbf{v}}_t}+\epsilon} + \lambda \theta_{t-1}\right),
\end{equation}
where $\lambda$ is the weight decay coefficient. Adam/AdamW is almost employed as the default choice across both academic research and industrial production for recommendation models.

\subsection{Muon Optimizer}
Muon~\cite{liu2025muon,jordan2024muon} is designed to optimize \textbf{2D matrix parameters} (\eg, linear layer weights).
For a matrix parameter $\mathbf{W}\in\mathbb{R}^{n\times m}$ with gradient $\mathbf{G}_t=\nabla_{\mathbf{W}}\mathcal{L}_t$,
Muon is first formed in an SGD-momentum-style update~\cite{amari1993backpropagation}:
\begin{equation}
\mathbf{M}_t = \mu\mathbf{M}_{t-1} + \mathbf{G}_t,
\end{equation}
where $\mu$ is the momentum coefficient. Then a Newton--Schulz iteration~\cite{jordan2024muon} is applied to approximately orthogonalize the momentum:
\begin{equation}
\mathbf{O}_t = \operatorname{NS}(\mathbf{M}_t)\ \approx\ \operatorname{Ortho}(\mathbf{M}_t),
\end{equation}
where $\operatorname{NS}(\cdot)$ denotes the Newton–Schulz iteration, and  $\operatorname{Ortho}(\cdot)$ denotes the nearest semi-orthogonal matrix in Frobenius norm.
Equivalently, if $\mathbf{M}_t=\mathbf{U}\mathbf{\Sigma}\mathbf{V}^\top$ is the Singular Value Decomposition (SVD), then $\mathrm{Ortho}(\mathbf{M}_t)=\mathbf{U}\mathbf{V}^\top$, which is equivalent to $(\mathbf{M}_t\mathbf{M}_t^\top)^{-1/2}\mathbf{M}_t$ when $n \le m$. Finally, Muon updates the parameter by:
\begin{equation}
\mathbf{W}_t = \mathbf{W}_{t-1} - \eta\,\mathbf{O}_t,
\end{equation}
and a decoupled weight decay term can be added analogously:
\begin{equation}
\mathbf{W}_t = \mathbf{W}_{t-1} - \eta\left(\mathbf{O}_t + \lambda \mathbf{W}_{t-1}\right).
\end{equation}


\subsection{MuonRec}
\label{sec:MuonRec}

In this paper, we propose \textbf{\textit{MuonRec}}, which replaces Adam/AdamW with Muon for \textit{matrix-valued hidden-layer weights} in recommendation models. Concretely, we partition parameters into two groups and adopt a hybrid strategy to optimize them respectively:
\begin{itemize}[leftmargin=10pt]
    \item \textbf{Muon group:} all 2D weight matrices inside Transformer and sequence modeling blocks (\eg, attention projections and Multi-Layer Perceptron weights).
    \item \textbf{Adam group:} 1D parameters (bias/LayerNorm), embeddings and output layers (\eg, SID embedding and softmax heads in generative recommendation models).
\end{itemize}
\textit{Why Muon works in recommendation.} While a theoretical consensus on Muon is still emerging, we provide two intuitions on its efficacy:
\begin{itemize}[leftmargin=10pt]
    \item \textbf{Balanced update directions.}
In Transformer-based recommenders, per-parameter optimizers (\eg, Adam) can bias updates toward a low-rank subspace. Muon orthogonalizes the momentum to recover suppressed gradient directions, reducing dominance by a few eigenvectors and improving optimization stability and representation diversity.

    \item \textbf{Matrix-aware geometry for weight operators.}
    Muon treats 2D parameters as linear operators and performs an approximate operator-norm-aware step via the orthogonalized momentum,
    which can be more aligned with how weight matrices act on hidden representations than coordinate-wise adaptive scaling.
\end{itemize}
Our work is an empirical study of Muon in recommendation, demonstrating its effectiveness and efficiency in recommendation model training. We hope MuonRec can motivate broader exploration of optimizer design for large-scale recommendation.

\begin{table*}[t]
\centering
\vspace{-10pt}
\caption{
Performance comparison with Adam on \textit{Industrial} and \textit{Official} datasets.
For MiniOneRec models, we report the \textbf{converged training step} during the SFT+RL phase.
Best results are in \textbf{bold}. N@$K$: NDCG@$K$.  R@$K$: Recall@$K$.
}
\vspace{-7pt}
\label{tab:main_results_merged}
\resizebox{\textwidth}{!}{%
\setlength{\tabcolsep}{3.5pt} 
\renewcommand{\arraystretch}{0.95} 
\definecolor{lightgray}{gray}{0.9}
\begin{tabular}{lc l cccccccc c cccccccc}
\toprule
\multirow{2}{*}{\textbf{Type}} & \multirow{2}{*}{\textbf{Model}} & \multirow{2}{*}{\textbf{Opt.}} & \multicolumn{8}{c}{\textbf{Industrial Dataset}} & \multicolumn{8}{c}{\textbf{Official Dataset}} \\
\cmidrule(lr){4-11} \cmidrule(lr){12-19}
 &  &  & Step $\downarrow$ & R@1 & R@3 & R@5 & R@10 & N@3 & N@5 & N@10 & Step $\downarrow$ & R@1 & R@3 & R@5 & R@10 & N@3 & N@5 & N@10 \\
\midrule

\multirow{9}{*}{Trad.} 
& \multirow{3}{*}{GRU4Rec} 
  & Adam & 110 & 0.0516 & 0.0746 & 0.0863 & 0.1054 & 0.0651 & 0.0699 & 0.0761 
          & 142 & 0.0506 & 0.0816 & 0.0960 & 0.1210 & 0.0687 & 0.0746 & 0.0826 \\
& & Muon & \textbf{44} & \textbf{0.0598} & \textbf{0.0869} & \textbf{0.0971} & \textbf{0.1163} & \textbf{0.0753} & \textbf{0.0794} & \textbf{0.0855} 
          & \textbf{86} & \textbf{0.0705} & \textbf{0.1007} & \textbf{0.1099} & \textbf{0.1311} & \textbf{0.0881} & \textbf{0.0919} & \textbf{0.0987} \\
& & \cellcolor{lightgray}\textit{Improv.} & \cellcolor{lightgray}60.0\% & \cellcolor{lightgray}15.9\% & \cellcolor{lightgray}16.5\% & \cellcolor{lightgray}12.5\% & \cellcolor{lightgray}10.3\% & \cellcolor{lightgray}15.7\% & \cellcolor{lightgray}13.6\% & \cellcolor{lightgray}12.4\% 
          & \cellcolor{lightgray}39.4\% & \cellcolor{lightgray}39.3\% & \cellcolor{lightgray}23.4\% & \cellcolor{lightgray}14.5\% & \cellcolor{lightgray}8.3\% & \cellcolor{lightgray}28.2\% & \cellcolor{lightgray}23.2\% & \cellcolor{lightgray}19.5\% \\
\cmidrule{2-19}

& \multirow{3}{*}{SASRec} 
  & Adam & 117 & \textbf{0.0576} & 0.0805 & 0.0931 & \textbf{0.1141} & 0.0713 & 0.0764 & \textbf{0.0831} 
          & 153 & 0.0717 & 0.0935 & 0.1060 & 0.1262 & 0.0847 & 0.0898 & 0.0963 \\
& & Muon & \textbf{103} & 0.0569 & \textbf{0.0810} & \textbf{0.0944} & 0.1127 & \textbf{0.0713} & \textbf{0.0769} & 0.0827 
          & \textbf{117} & \textbf{0.0732} & \textbf{0.0962} & \textbf{0.1087} & \textbf{0.1270} & \textbf{0.0866} & \textbf{0.0918} & \textbf{0.0977} \\
& & \cellcolor{lightgray}\textit{Improv.} & \cellcolor{lightgray}12.0\% & \cellcolor{lightgray}-1.2\% & \cellcolor{lightgray}0.6\% & \cellcolor{lightgray}1.4\% & \cellcolor{lightgray}-1.2\% & \cellcolor{lightgray}0.0\% & \cellcolor{lightgray}0.7\% & \cellcolor{lightgray}-0.5\% 
          & \cellcolor{lightgray}23.5\% & \cellcolor{lightgray}2.1\% & \cellcolor{lightgray}2.9\% & \cellcolor{lightgray}2.5\% & \cellcolor{lightgray}0.6\% & \cellcolor{lightgray}2.2\% & \cellcolor{lightgray}2.2\% & \cellcolor{lightgray}1.5\% \\
\cmidrule{2-19}

& \multirow{3}{*}{Caser} 
  & Adam & 192 & \textbf{0.0406} & 0.0457 & 0.0483 & 0.0571 & 0.0436 & 0.0447 & 0.0475 
          & 75 & 0.0360 & 0.0584 & 0.0754 & 0.0997 & 0.0487 & 0.0557 & 0.0635 \\
& & Muon & \textbf{145} & 0.0386 & \textbf{0.0629} & \textbf{0.0715} & \textbf{0.0898} & \textbf{0.0529} & \textbf{0.0565} & \textbf{0.0624} 
          & \textbf{44} & \textbf{0.0409} & \textbf{0.0771} & \textbf{0.0896} & \textbf{0.1106} & \textbf{0.0618} & \textbf{0.0669} & \textbf{0.0736} \\
& & \cellcolor{lightgray}\textit{Improv.} & \cellcolor{lightgray}24.5\% & \cellcolor{lightgray}-4.9\% & \cellcolor{lightgray}37.6\% & \cellcolor{lightgray}48.0\% & \cellcolor{lightgray}57.3\% & \cellcolor{lightgray}21.3\% & \cellcolor{lightgray}26.4\% & \cellcolor{lightgray}31.4\% 
          & \cellcolor{lightgray}41.3\% & \cellcolor{lightgray}13.6\% & \cellcolor{lightgray}32.0\% & \cellcolor{lightgray}18.8\% & \cellcolor{lightgray}10.9\% & \cellcolor{lightgray}26.9\% & \cellcolor{lightgray}20.1\% & \cellcolor{lightgray}15.9\% \\

\midrule

\multirow{12}{*}{Gen.} 
& \multirow{3}{*}{TIGER} 
  & Adam & 3600 & 0.0711 & 0.0865 & 0.0940 & \textbf{0.1114} & 0.0801 & 0.0831 & 0.0889 
          & 3900 & 0.0752 & 0.1008 & 0.1103 & 0.1226 & 0.0908 & 0.0947 & 0.0987 \\
& & Muon & \textbf{1200} & \textbf{0.0731} & \textbf{0.0870} & \textbf{0.0965} & 0.1094 & \textbf{0.0814} & \textbf{0.0853} & \textbf{0.0895} 
          & \textbf{1200} & \textbf{0.1270} & \textbf{0.1549} & \textbf{0.1655} & \textbf{0.1766} & \textbf{0.1436} & \textbf{0.1480} & \textbf{0.1516} \\
& & \cellcolor{lightgray}\textit{Improv.} & \cellcolor{lightgray}66.7\% & \cellcolor{lightgray}2.8\% & \cellcolor{lightgray}0.6\% & \cellcolor{lightgray}2.7\% & \cellcolor{lightgray}-1.8\% & \cellcolor{lightgray}1.6\% & \cellcolor{lightgray}2.6\% & \cellcolor{lightgray}0.7\% 
          & \cellcolor{lightgray}69.2\% & \cellcolor{lightgray}68.9\% & \cellcolor{lightgray}53.7\% & \cellcolor{lightgray}50.0\% & \cellcolor{lightgray}44.0\% & \cellcolor{lightgray}58.1\% & \cellcolor{lightgray}56.3\% & \cellcolor{lightgray}53.6\% \\
\cmidrule{2-19}

& \multirow{3}{*}{\shortstack{MiniOneRec\\(0.5B)}} 
  & Adam & 3675 & 0.0730 & 0.1013 & 0.1147 & 0.1293 & 0.0894 & 0.0949 & 0.0996 
          & 3700 & 0.0861 & 0.1145 & 0.1270 & 0.1406 & 0.1028 & 0.1079 & 0.1123 \\
& & Muon & \textbf{3345} & \textbf{0.0768} & \textbf{0.1026} & \textbf{0.1200} & \textbf{0.1403} & \textbf{0.0916} & \textbf{0.0988} & \textbf{0.1054} 
          & \textbf{2710} & \textbf{0.0906} & \textbf{0.1235} & \textbf{0.1356} & \textbf{0.1560} & \textbf{0.1099} & \textbf{0.1150} & \textbf{0.1216} \\
& & \cellcolor{lightgray}\textit{Improv.} & \cellcolor{lightgray}9.0\% & \cellcolor{lightgray}5.2\% & \cellcolor{lightgray}1.3\% & \cellcolor{lightgray}4.6\% & \cellcolor{lightgray}8.5\% & \cellcolor{lightgray}2.5\% & \cellcolor{lightgray}4.1\% & \cellcolor{lightgray}5.8\% 
          & \cellcolor{lightgray}26.8\% & \cellcolor{lightgray}5.2\% & \cellcolor{lightgray}7.9\% & \cellcolor{lightgray}6.8\% & \cellcolor{lightgray}11.0\% & \cellcolor{lightgray}6.9\% & \cellcolor{lightgray}6.6\% & \cellcolor{lightgray}8.3\% \\
\cmidrule{2-19}

& \multirow{3}{*}{\shortstack{MiniOneRec\\(1.5B)}} 
  & Adam & 3345 & 0.0781 & 0.1046 & 0.1160 & 0.1372 & 0.0933 & 0.0980 & 0.1048 
          & 3700 & 0.0933 & 0.1210 & 0.1313 & 0.1476 & 0.1097 & 0.1140 & 0.1192 \\
& & Muon & \textbf{2355} & \textbf{0.0788} & \textbf{0.1066} & \textbf{0.1253} & \textbf{0.1460} & \textbf{0.0949} & \textbf{0.1026} & \textbf{0.1092} 
          & \textbf{2710} & \textbf{0.0933} & \textbf{0.1264} & \textbf{0.1383} & \textbf{0.1560} & \textbf{0.1125} & \textbf{0.1174} & \textbf{0.1231} \\
& & \cellcolor{lightgray}\textit{Improv.} & \cellcolor{lightgray}29.6\% & \cellcolor{lightgray}0.9\% & \cellcolor{lightgray}1.9\% & \cellcolor{lightgray}8.0\% & \cellcolor{lightgray}6.4\% & \cellcolor{lightgray}1.7\% & \cellcolor{lightgray}4.7\% & \cellcolor{lightgray}4.2\% 
          & \cellcolor{lightgray}26.8\% & \cellcolor{lightgray}0.0\% & \cellcolor{lightgray}4.5\% & \cellcolor{lightgray}5.3\% & \cellcolor{lightgray}5.7\% & \cellcolor{lightgray}2.6\% & \cellcolor{lightgray}3.0\% & \cellcolor{lightgray}3.3\% \\
\cmidrule{2-19}

& \multirow{3}{*}{\shortstack{MiniOneRec\\(3.0B)}} 
  & Adam & 2790 & 0.0737 & 0.0993 & 0.1088 & 0.1260 & 0.0883 & 0.0922 & 0.0977 
          & 2870 & 0.0816 & 0.1151 & 0.1243 & 0.1393 & 0.1010 & 0.1048 & 0.1096 \\
& & Muon & \textbf{2430} & \textbf{0.0777} & \textbf{0.1032} & \textbf{0.1205} & \textbf{0.1423} & \textbf{0.0920} & \textbf{0.0991} & \textbf{0.1062} 
          & \textbf{2540} & \textbf{0.0878} & \textbf{0.1219} & \textbf{0.1348} & \textbf{0.1496} & \textbf{0.1077} & \textbf{0.1131} & \textbf{0.1178} \\
& & \cellcolor{lightgray}\textit{Improv.} & \cellcolor{lightgray}12.9\% & \cellcolor{lightgray}5.4\% & \cellcolor{lightgray}3.9\% & \cellcolor{lightgray}10.8\% & \cellcolor{lightgray}12.9\% & \cellcolor{lightgray}4.2\% & \cellcolor{lightgray}7.5\% & \cellcolor{lightgray}8.7\% 
          & \cellcolor{lightgray}11.5\% & \cellcolor{lightgray}7.6\% & \cellcolor{lightgray}5.9\% & \cellcolor{lightgray}8.4\% & \cellcolor{lightgray}7.4\% & \cellcolor{lightgray}6.6\% & \cellcolor{lightgray}7.9\% & \cellcolor{lightgray}7.5\% \\

\bottomrule
\end{tabular}%
}
\end{table*}

\section{Experiments}
\label{sec:experiment}


\subsection{Experiment Setup}

\subsubsection{Datasets.}
Our experiments are conducted on two public datasets from the Amazon Review collection~\cite{ni2019justifying}: \textit{Industrial and Scientific} and \textit{Office Products}. Following previous works~\cite{kong2025minionerecopensourceframeworkscaling, hou2024bridging, zhai2024actions}, we apply a 5-core filtering and a chronological leave-one-out strategy to preprocess the data. After preprocessing, the \textit{Industrial} dataset comprises 1,987 users, 3,686 items and 13,555 interactions, while \textit{Office} contains 2,024 users, 3,459 items and 13,929 interactions.


\subsubsection{Evaluated Models.}

We compare MuonRec with Adam~\cite{kingma2014adam} on models spanning two distinct paradigms, \ie, \textit{traditional sequential recommendation} and \textit{generative recommendation}. For the traditional paradigm, we select \textbf{GRU4Rec}~\cite{hidasi2015session}, \textbf{Caser}~\cite{tang2018personalized}, and \textbf{SASRec}~\cite{kang2018self}. For the generative paradigm, we choose two representatives, \ie, \textbf{TIGER}~\cite{rajput2024recommender} and \textbf{MiniOneRec}~\cite{kong2025minionerecopensourceframeworkscaling}.


\subsubsection{Evaluation Metrics.}
 Following standard practice~\cite{kong2025minionerecopensourceframeworkscaling, chen2024hllm, he2015trirank}, we employ two metrics, \ie, Recall@$K$ and NDCG@$K$ (Normalized Discounted Cumulative Gain), to evaluate \textit{top-$K$ recommendation performance}. We report the experimental results with $K \in \{1, 3, 5, 10\}$. For both metrics, higher values indicate better performance.

To evaluate the \textit{convergence speed}, we report the number of training steps required to reach convergence, under a fixed batch size and a unified early-stopping criterion across different models. Fewer steps mean faster convergence.


\subsubsection{Implementation Details.} 
For all traditional models, we standardize the hidden dimension to 64 and the training batch size to 256. For TIGER, we employ a Transformer backbone consisting of 4 encoder and 4 decoder layers with a hidden dimension of 128, trained with a global batch size of 256. The training of MiniOneRec comprises two phases, utilizing \textit{Qwen2.5-Instruct}~\cite{yang2025qwen3} as the backbone model. The Supervised Fine-Tuning (SFT) phase is conducted with a global batch size of 1,024 for up to 10 epochs, incorporating an early stopping patience of one epoch. Subsequently, the Reinforcement Learning (RL) phase employs the GRPO algorithm with a batch size of 512 and a rollout number of 16.

As noted in Section~\ref{sec:MuonRec}, Muon optimizes 2D internal weight matrices, while Adam handles vector-wise parameters. To ensure a rigorous evaluation, we enforce a strict two-stage tuning protocol. 
First, we tune the entire model using Adam scanning learning rates from $\{10^{-5}, 3 \times 10^{-5}, 10^{-4}, 3 \times 10^{-4}, 10^{-3}, 3 \times 10^{-3}\}$ and weight decay from $\{10^{-5}, 10^{-4}, 10^{-3}, 10^{-2}\}$. 
Second, while fixing the vector-wise parameters at their optimal values, we exclusively tune the Muon-optimized group, scanning learning rates from $\{10^{-5}, 3 \times 10^{-5}, 10^{-4}, 3 \times 10^{-4}, 10^{-3}, 3 \times 10^{-3}, 10^{-2}\}$ and weight decay from $\{10^{-5}, 5 \times 10^{-5}, 10^{-4}, 5 \times 10^{-4}, 10^{-3}, 5 \times 10^{-3}\}$. 
This protocol effectively isolates the performance gains to Muon's orthogonalized updates. All experiments are conducted on NVIDIA H100 GPUs.


\subsection{Training Efficiency and Effectiveness}
Table~\ref{tab:main_results_merged} and Figure~\ref{fig:tiger_dynamics} summarize the performance of MuonRec.
\begin{itemize}[leftmargin=10pt]
\item \textbf{Convergence speed.} MuonRec consistently accelerates convergence for both traditional and generative recommenders, yielding an average reduction of 32.4\% in training steps across all tested scenarios. For \textit{traditional} models, Muon reduces the training steps by an average of 33.5\% on two datasets. For \textit{generative} models, the speedup is more pronounced. TIGER converges roughly 3$\times$ faster on Industrial. Figure~\ref{fig:tiger_dynamics} further shows a steeper loss descent, suggesting that Muon's orthogonalized updates enable more efficient optimization than Adam.

\item \textbf{Recommendation performance.}
Beyond acceleration, MuonRec also boosts ranking quality across architectures. On \textit{traditional} models, Muon improves Recall and NDCG by 15.1\% and 14.5\% on average across both datasets. On \textit{Official} for \textit{generative} recommendation, TIGER sees much larger gains: +54.2\% Recall and +56.0\% NDCG. We attribute this to Muon strengthening minority gradient directions and avoiding collapse into a few dominant eigenvectors. As an implicit spectral regularizer, it encourages a more uniform singular-value spectrum, which may reduce index collapse in quantization-based generative recommenders~\cite{li2025survey, zhou2025openonerectechnicalreport}.

\item \textbf{Scalability on large generative models.} MuonRec remains effective when scaling up on MiniOneRec, improving training efficiency without sacrificing accuracy. For MiniOneRec-1.5B, Muon reduces the converged training steps by over \textbf{25\%} on both datasets, while maintaining consistently better ranking performance than Adam. Moreover, larger MiniOneRec models generally deliver stronger results than smaller ones (\eg, 1.5B vs.\ 0.5B), indicating that Muon supports effective scaling in large-model training. Notably, the 3.0B variant underperforms 1.5B under both optimizers, which we attribute to limited training data that exacerbates overfitting. This hypothesis can be further examined by enlarging the training set or strengthening regularization.
\end{itemize}




\label{sec:main_results}

\begin{table}[t]
\vspace{-2.4pt}
\centering
\caption{
Optimizer combinations for SFT and RL stages. 
\textit{Step} denotes the total converged training steps of both stages. 
\textbf{Bold}: best results; \underline{underlined}: second best results.
}
\label{tab:combinations_ablation_refactored}
\vspace{-8pt}
\resizebox{0.47\textwidth}{!}{
\renewcommand{\arraystretch}{1.1} 
\setlength{\tabcolsep}{3.5pt}      
\resizebox{\linewidth}{!}{
\begin{tabular}{cc c cccc ccc} 
\toprule
\multicolumn{2}{c}{\textbf{Opt.}} & \multirow{2}{*}{\textbf{Step} $\downarrow$} & \multicolumn{4}{c}{\textbf{Recall}} & \multicolumn{3}{c}{\textbf{NDCG } } \\
\cmidrule(lr){1-2} \cmidrule(lr){4-7} \cmidrule(lr){8-10}
SFT & RL & & R@1 & R@3 & R@5 & R@10 & N@3 & N@5 & N@10 \\ 
\midrule
\multicolumn{10}{c}{\textbf{\textit{Dataset: Industrial}}} \\ 
\midrule
Adam & Adam & 3345 & 0.0781 & 0.1046 & 0.1160 & 0.1372 & 0.0933 & 0.0980 & 0.1048 \\
Adam & Muon & 3015 & 0.0748 & 0.1026 & 0.1176 & 0.1410 & 0.0907 & 0.0968 & 0.1045 \\
Muon & Adam & \underline{2685} & \textbf{0.0796} & \textbf{0.1079} & \underline{0.1231} & \underline{0.1423} & \textbf{0.0959} & \underline{0.1021} & \underline{0.1083} \\
Muon & Muon & \textbf{2355} & \underline{0.0788} & \underline{0.1066} & \textbf{0.1253} & \textbf{0.1460} & \underline{0.0949} & \textbf{0.1026} & \textbf{0.1092} \\ 
\midrule
\multicolumn{10}{c}{\textbf{\textit{Dataset: Official}}} \\ 
\midrule
Adam & Adam & 3700 & \underline{0.0933} & 0.1210 & 0.1313 & 0.1476 & 0.1097 & 0.1140 & 0.1192 \\
Adam & Muon & 3370 & 0.0894 & 0.1212 & 0.1332 & 0.1471 & 0.1082 & 0.1130 & 0.1176 \\
Muon & Adam & \underline{3040} & \textbf{0.0972} & \underline{0.1229} & \underline{0.1344} & \underline{0.1510} & \underline{0.1124} & \underline{0.1171} & \underline{0.1225} \\
Muon & Muon & \textbf{2710} & \underline{0.0933} & \textbf{0.1264} & \textbf{0.1383} & \textbf{0.1560} & \textbf{0.1125} & \textbf{0.1174} & \textbf{0.1231} \\ 
\bottomrule
\end{tabular}}
}
\vspace{-13pt}
\end{table}

\subsection{In-depth Analysis}
\subsubsection{Impact of Optimizer Strategy.}
In this part, we investigate how various optimizer configurations affect the multi-stage training process (\ie, SFT and RL) of GenRec models. We report the results on MiniOneRec-1.5B in Table~\ref{tab:combinations_ablation_refactored}.

Empirically, applying Muon in both stages is the most effective strategy. It achieves the best results on most metrics while requiring the fewest training steps. Using MuonRec for SFT and Adam for RL is typically second-best, indicating that configurations \emph{starting} with Muon consistently outperform those starting with Adam. For example, on \textit{Office}, Muon in SFT clearly outperforms introducing Muon only during RL.
We attribute this gap to \textit{optimization-trajectory consistency} with the base model. Because the backbone \textit{Qwen2.5-Instruct}~\cite{yang2025qwen3} is trained with Adam, switching to Muon’s orthogonalized updates constitutes a substantial change in optimization dynamics. Making this switch during SFT, when data is abundant, gives the model enough updates to adapt its weight space to the new geometric constraints. In contrast, introducing Muon only in RL, which uses fewer steps and tighter objectives, can disrupt optimization momentum and yield weaker alignment.

\subsubsection{Hyperparameter Study.} 
\begin{figure}[t]
  \centering
  \includegraphics[width=\linewidth]{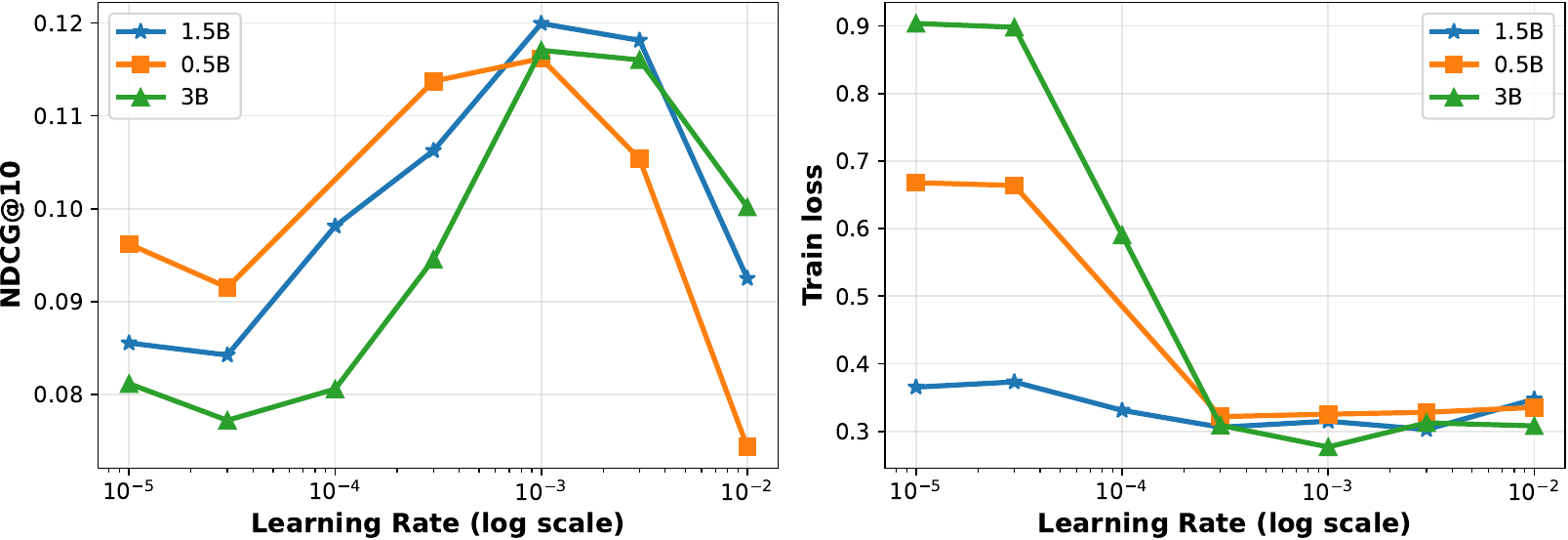}
  \vspace{-0.6cm}
  \caption{Impact of LR on NDCG@10 and training loss.}
  
  \vspace{-0.5cm}
  \label{sft}
\end{figure}

We investigate the impact of learning rate $\eta$ on the \textit{Office} dataset across varying model scales (0.5B, 1.5B, and 3.0B). 
Figure~\ref{sft} (left) reveals that performance follows a consistent bell-shaped trend, peaking at an optimal regime around $\eta \approx 10^{-2}$. Unlike Adam, Muon benefits from a more aggressive regime due to the implicit spectral regularization of the Newton-Schulz iteration. Accordingly, we recommend initializing $\eta$ within $[5 \times 10^{-3}, 5 \times 10^{-2}]$ for optimal stability.

Crucially, this optimal regime exhibits scale invariance, which is consistent with results in previous work~\cite{liu2025muon}: the peak remains anchored at  the same log scale (\eg, $\eta \approx 10^{-2}$) regardless of model size. This consistency enables efficient hyperparameter tuning on smaller proxy models before transferring to billion-scale counterparts, reducing computational costs. As shown in Figure~\ref{sft} (right), the lowest training loss aligns perfectly with peak NDCG at $\eta=10^{-2}$, whereas excessive rates ($\eta \ge 10^{-1}$) lead to sharp divergence.

\section{Related Work}
\label{sec: related work}

Adaptive per-parameter optimizers, particularly Adam~\cite{kingma2014adam}, remain the default optimizer for training recommendation models~\cite{xi2024towards, lin2024rella, liu2024mamba4rec, lin2024clickprompt, shan2025automatic, zhu2024lifelong}. However, as recommenders evolve toward deeper Transformer-based architectures, they increasingly suffer from representation collapse that per-parameter optimizers struggle to address.

Muon is as a matrix-structured optimizer that orthogonalizes momentum updates for 2D weights with Newton-Schulz iterations, initially to accelerate small-scale language model training~\cite{jordan2024muon}. To facilitate large-scale deployment, Moonlight~\cite{liu2025muon} and Flash-Muon~\cite{lin2025flash} improve communication efficiency and hardware acceleration. Further extensions address stability and scalability through diverse mechanisms: NorMuon~\cite{li2025normuon} incorporates neuron-wise adaptive scaling and row-wise normalization, MuonClip~\cite{team2025kimi} introduces QK-clip style rescaling to prevent attention explosion. Riemannion~\cite{riemannionparametrization} generalizes the framework to Riemannian optimization for LoRA. On the theory side, variants like Muon-VR2~\cite{chang2025convergence} provide enhanced convergence guarantees through variance reduction.

However, the feasibility of Muon in recommendation model training remains an open research question, and this work serves as the first empirical study of Muon in recommendation, demonstrating better training efficiency and efficacy than widely used Adam.

\section{Conclusion}

In this paper, we revisit optimizer design for modern recommender systems and introduced \textbf{MuonRec}, the first framework that adapts the Muon optimizer to RecSys training. By applying orthogonalized momentum updates for $2$D weight matrices via Newton--Schulz iteration, MuonRec improves optimization efficiency and yields better-conditioned updates. Extensive experiments on both sequential and generative recommenders show that MuonRec consistently outperforms Adam, achieving faster convergence and stronger final ranking quality under comparable compute budgets. On average, MuonRec reduces the number of converged training steps by \textbf{32.4\%} and improves NDCG@10 by \textbf{12.6\%} across all settings. Overall, our results position Muon as a promising optimizer alternative for modern large-scale RecSys training, and motivate future work on further optimizations for Muon in recommendation.


\bibliographystyle{ACM-Reference-Format}
\bibliography{acmart}

\newpage

\appendix


\end{document}